\documentclass[aps,prb,twocolumn,groupedaddress,floatfix,showpacs]{revtex4-1}
\usepackage{graphics}
\usepackage{subfigure}
\usepackage{epsfig}
\usepackage{dcolumn}
\usepackage{bm}
\usepackage{overpic} 

\usepackage{amsmath}
\usepackage[utf8]{inputenc}
\usepackage{pgfplots}
\pgfplotsset{compat=1.18}
\usepackage{siunitx}

\begin{document}

\title{Decoupling Composition and Band Gap in $\kappa$-Ga$_2$O$_3$ Heterostructures via STEM-EELS}

\author{Annett Th\o gersen}
\affiliation{SINTEF Industry, Materials Physics, Forskningsveien 1, 0373 Oslo, Norway}

\author{Georg Muntingh}
\affiliation{SINTEF Digital, Forskningsveien 1, 0373 Oslo, Norway}

\author{Lasse Vines}
\affiliation{Department of Physics, Centre for Materials Science and Nanotechnology, University of Oslo, P.O. Box 1048 Blindern, N-0316 Oslo, Norway}

\author{Øystein Prytz}
\affiliation{Department of Physics, Centre for Materials Science and Nanotechnology, University of Oslo, P.O. Box 1048 Blindern, N-0316 Oslo, Norway}

\author{Max Knei\ss}
\author{Marius Grundmann}
\author{Holger von Wenckstern}
\affiliation{University of Leipzig, Felix-Bloch-Institut für Festkörperphysik, Linnéstrasse 5, 04103 Leipzig, Germany }

\author{Ingvild J. T. Jensen}
\affiliation{SINTEF Industry, Materials Physics, Forskningsveien 1, 0373 Oslo, Norway}

\date{\today}

\begin{abstract}
High-resolution mapping of electronic properties at oxide heterointerfaces remains challenging due to probe delocalization and overlapping signals. In this work, we employ monochromated, probe-corrected scanning transmission electron microscopy combined with electron energy-loss spectroscopy (STEM-EELS) to resolve band gap variations across $\kappa$-Ga$_2$O$_3$-based multilayers with nanometer-scale precision. A custom automated quantitative-based EELS analysis framework enabled automated band gap fitting and visualization, ensuring reproducibility and high spatial resolution. By optimizing acquisition parameters and quantifying inelastic delocalization, we demonstrate reliable extraction of band gap excitations from layers only a few nanometers thick. For heterostructures grown on ITO templates, strain at defect-free interfaces induces a gradual band gap transition from $5.08~\mathrm{eV}$ to $4.28~\mathrm{eV}$ over $\sim 10~\mathrm{nm}$, despite an abrupt compositional change. In contrast, ZnO-based templates introduce structural defects that relieve strain, yielding band gaps consistent with composition. These results establish STEM-EELS as a powerful tool for nanoscale electronic characterization and highlight the dominant role of interfacial strain over composition in governing local band structure.

\end{abstract}

\maketitle

\section{Introduction}

Understanding electronic structure at oxide heterointerfaces is critical for designing next-generation functional materials. In complex oxides, local variations in band gap influence charge transport, interface dipoles, and defect formation, which in turn affect device performance and reliability. Gallium oxide (Ga$_2$O$_3$) has emerged as a key wide band gap semiconductor due to its ultra-wide band gap (4.5--5.3~eV), high breakdown strength, and multiple polymorphs ($\alpha$, $\beta$, $\gamma$, $\delta$, and $\kappa$), each offering distinct structural and electronic characteristics.\cite{doi:10.1021/ja01123a039} Among these, the $\kappa$-phase is particularly attractive because of its polar orthorhombic \textit{Pna2$_1$} structure, which enables spontaneous polarization and the formation of two-dimensional electron gases (2DEGs) at heterointerfaces.\cite{ref:cora2017, Nikolaev_2020}

Alloying $\kappa$-Ga$_2$O$_3$ with Al or In provides a powerful means of tuning its band gap and carrier concentration, enabling tailored properties for high-power and high-frequency applications.\cite{Hassa_2021} The band gap dependence on composition for $\kappa$-(Al$_x$Ga$_{1-x}$)$_2$O$_3$ and $\kappa$-(In$_y$Ga$_{1-y}$)$_2$O$_3$ can be approximated as $E_g = 4.91 + 2.10x$ and $E_g = 4.90 - 1.95y$, respectively, with miscibility limits of $x \leq 0.35$ and $y \leq 0.62$. These relationships predict band gaps spanning 4.2--6.2~eV, but experimental observations often deviate from these values, particularly near heterointerfaces.\cite{Thogersen:soon}

A central question is whether band gap transitions across heterointerfaces mirror compositional changes or are modified by strain, defects, or measurement artifacts. Interfaces that appear compositionally sharp frequently exhibit gradual band gap variations, suggesting a possible decoupling between composition and electronic structure. Resolving this requires a technique capable of mapping band gap excitations with nanometer-scale spatial resolution while accounting for artifacts such as inelastic delocalization.

Electron energy-loss spectroscopy (EELS) in the scanning transmission electron microscope (STEM) is uniquely suited for probing local electronic structure, but its interpretation in layered systems is complicated by spatial resolution limits.\cite{EGERTON2007575, STOGERPOLLACH201498} Inelastic delocalization, arising from long-range Coulomb interactions, extends the effective interaction volume beyond the probe size, potentially blending signals from adjacent layers. This effect, combined with beam broadening and chromatic aberrations, can obscure abrupt electronic transitions. Quantifying and mitigating these artifacts is essential for reliable band gap measurements. While recent studies have demonstrated automated band gap mapping in STEM-EELS,\cite{Granerod2018} and spatially resolved band gap variations across Ga$_2$O$_3$ interfaces using valence-band EELS (VEELS),\cite{Chmielewski2022} these approaches often lack robust workflows for reproducible, high-throughput analysis and visualization.

In this work, we address this gap by combining monochromated, probe-corrected STEM-EELS with a custom automated quantitative-based analysis framework for automated band gap fitting and mapping. This approach enables reproducible extraction of band gap excitations from spectral images and line scans, even in layers only a few nanometers thick. We apply this methodology to $\kappa$-Ga$_2$O$_3$ heterostructures grown on two templates: indium tin oxide (ITO), which promotes defect-free but strained interfaces, and zinc oxide (ZnO), which introduces structural defects that may relieve strain. By comparing these architectures, we disentangle the roles of strain and defects in governing local band gap behavior and establish a workflow for nanoscale electronic characterization in complex oxides.

\section{Experimental}

The Ga$_2$O$_3$-based heterostructures were deposited using pulsed laser deposition (PLD) on sapphire substrates pre-coated \emph{in situ} with either an indium tin oxide (ITO) or zinc oxide (ZnO) buffer layer defining the growth template. A 248 nm KrF excimer laser (Coherent LPX Pro 305) was used to ablate ceramic oxide targets with an energy density of 2.6 J/cm$^2$ and a target-to-substrate distance of 90 mm. For binary and ternary compositions, either single-phase Ga$_2$O$_3$ targets or radially segmented targets composed of semicircular In$_2$O$_3$, Al$_2$O$_3$, and Ga$_2$O$_3$ were used. The method and details are described in a paper by Wenckstern et al. \cite{https://doi.org/10.1002/pssb.201900626} The targets contained tin, which is required to induce the formation of the $\kappa$-polymorph (surfactant-mediated growth, the tin stays primarily on the surface, only a small amount is incorporated into the layers). These targets were prepared from 99.999\% pure oxide powders, pressed and sintered at 1300\,$^\circ$C for 72 hours in air. In some cases, 0.1 wt.\% SiO$_2$ was added to enhance electrical conductivity. The deposition was carried out at substrate temperatures ranging from 450\,$^\circ$C to 650\,$^\circ$C and oxygen partial pressures between $3 \times 10^{-4}$ and $6 \times 10^{-3}$~mbar. 

Samples for transmission electron microscopy (TEM) were prepared using a focused ion beam (FIB) system (Helios G4 UX). The samples investigated in this work have a sample thickness range of 35--60~nm. The analyses were conducted on a DCOR Cs probe-corrected and monochromated FEI Titan G2 60-300 microscope with a 0.8 Å spatial resolution in STEM (at 300 kV), and about 2 Å at 60 kV. For band gap measurements, the microscope was operated at 60~kV, with a measured energy resolution of 0.15 eV, while high-resolution scanning TEM (STEM) imaging was performed at 300~kV. The system is equipped with a Gatan Quantum 965 electron energy loss spectroscopy (EELS) spectrometer and a Bruker Super-X energy-dispersive X-ray (EDX) detector, enabling comprehensive chemical and electronic structure characterization. Fast Fourier transforms (FFT) were used for indexing the structure and geometric phase analysis (GPA) with the FRWRtools plugin\cite{GPA} in Digital Micrograph (Gatan Inc) to evaluate strain.

EELS data were acquired both as spectral images (SI) and as line scans. The zero loss peak (ZLP) was extracted using a fitted log-tail method to ensure accurate energy calibration. Subsequent analysis of the SI datasets and individual spectra, including band gap fitting, was performed using a custom automated quantitative-based EELS analysis program developed in-house; more details are provided in the supplementary section.

\section{Results}

Two types of samples were investigated in this study. The first consists of a three-layer Ga$_2$O$_3$-based heterostructure grown on an ITO-coated sapphire substrate, as shown in Figure~\ref{figure:TEMall}A. The stack includes a pure Ga$_2$O$_3$ layer deposited directly on the ITO, followed by (In$_{0.18}$Ga$_{0.82}$)$_2$O$_3$, and capped with (Al$_{0.27}$Ga$_{0.73}$)$_2$O$_3$. A detailed structural and electronic analysis of this sample is presented in Section~\ref{section:ITO}. 

The second sample is a multilayered Ga$_2$O$_3$ heterostructure grown on a ZnO-coated sapphire substrate, designed to explore band gap variations across thin and compositionally distinct layers (Figure~\ref{figure:TEMall}B), as discussed in Section~\ref{section:zno}. High-resolution STEM imaging revealed several structural defects at the interfaces, such as stacking faults (Figure~\ref{figure:TEMall}D), likely caused by lattice mismatch or surface waviness inherited from the underlying ZnO.

\subsection{Strain analysis across the interface of the two samples}
\label{section:strain}

\begin{figure}
  \begin{center}
    \includegraphics[width=3.4in]{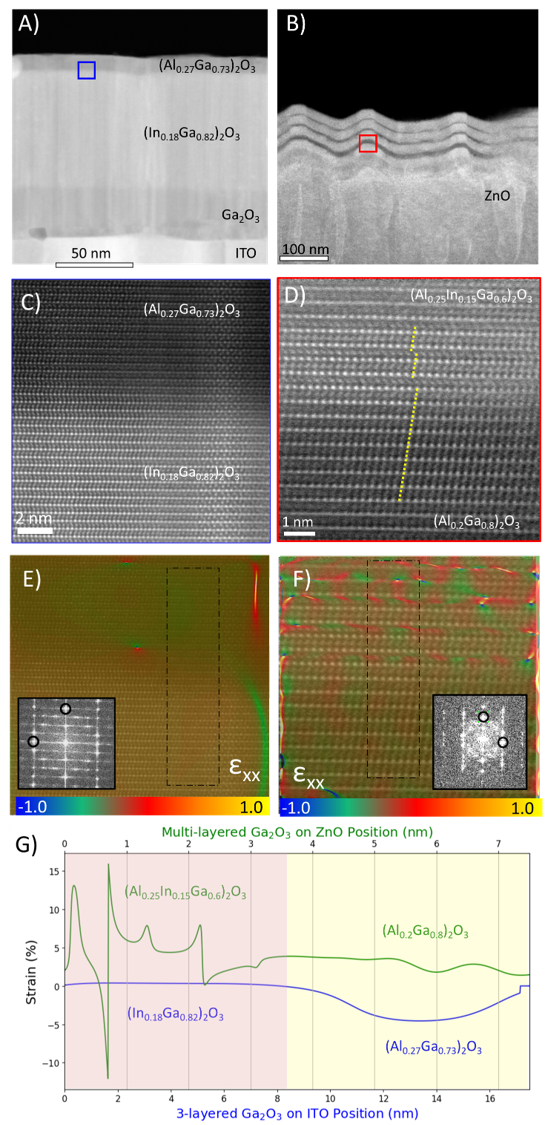}
          \caption{A) DF-STEM image of the 3-layered Ga$_2$O$_3$ on ITO substrate, B) DF-STEM image of doped, multi-layered Ga$_2$O$_3$ on ZnO, C) HR-STEM image of the first interface in the 3-layered Ga$_2$O$_3$ on ITO sample, D) HR-STEM image of an interface in the multi-layered Ga$_2$O$_3$ on ZnO, E) GPA strain map ($\varepsilon_{xx}$) of C), F) GPA strain map ($\varepsilon_{xx}$) of D), G) Strain profiles extracted from GPA for the two samples, showing variations across the heterointerfaces. Blue: 3-layered Ga$_2$O$_3$ on ITO; Green: multi-layered Ga$_2$O$_3$ on ZnO.}
    \label{figure:TEMall}
        \end{center}
\end{figure}

Local strain was analyzed using geometric phase analysis (GPA). Figures~\ref{figure:TEMall}E and~\ref{figure:TEMall}F show the $\varepsilon_{xx}$ strain maps corresponding to the interfaces highlighted in Figures~\ref{figure:TEMall}C and~\ref{figure:TEMall}D, respectively. Quantitative strain profiles extracted from these maps are presented in Figure~\ref{figure:TEMall}G, comparing the three-layered Ga$_2$O$_3$ on ITO (blue curve) and the multi-layered Ga$_2$O$_3$ on ZnO (green curve). The multi-layered Ga$_2$O$_3$ on ZnO exhibits pronounced local strain fluctuations at distinct defect sites, while almost no visible strain variation across the transition from the (Al$_{0.25}$In$_{0.15}$Ga$_{0.60}$)$_2$O$_3$ layer and into the (Al$_{0.20}$Ga$_{0.80}$)$_2$O$_3$ layer, and no areas with compressive strain. In contrast, the three-layered Ga$_2$O$_3$ on ITO displays does not show distinct defect sites in Figure~\ref{figure:TEMall}G, but shows a small defect in Figure~\ref{figure:TEMall}E. However, the strain curve shows a continuous trend from low strain to compressive strain (negative strain percentage) when moving from the (In$_{0.18}$Ga$_{0.82}$)$_2$O$_3$ layer to the (Al$_{0.27}$Ga$_{0.73}$)$_2$O$_3$ layer. this compressive strain can have an effect on the band gap of this material.

\subsection{Band gap variations across sharp interface in sample on ITO}
\label{section:ITO}

\begin{figure}
  \begin{center}
    \includegraphics[width=3.4in]{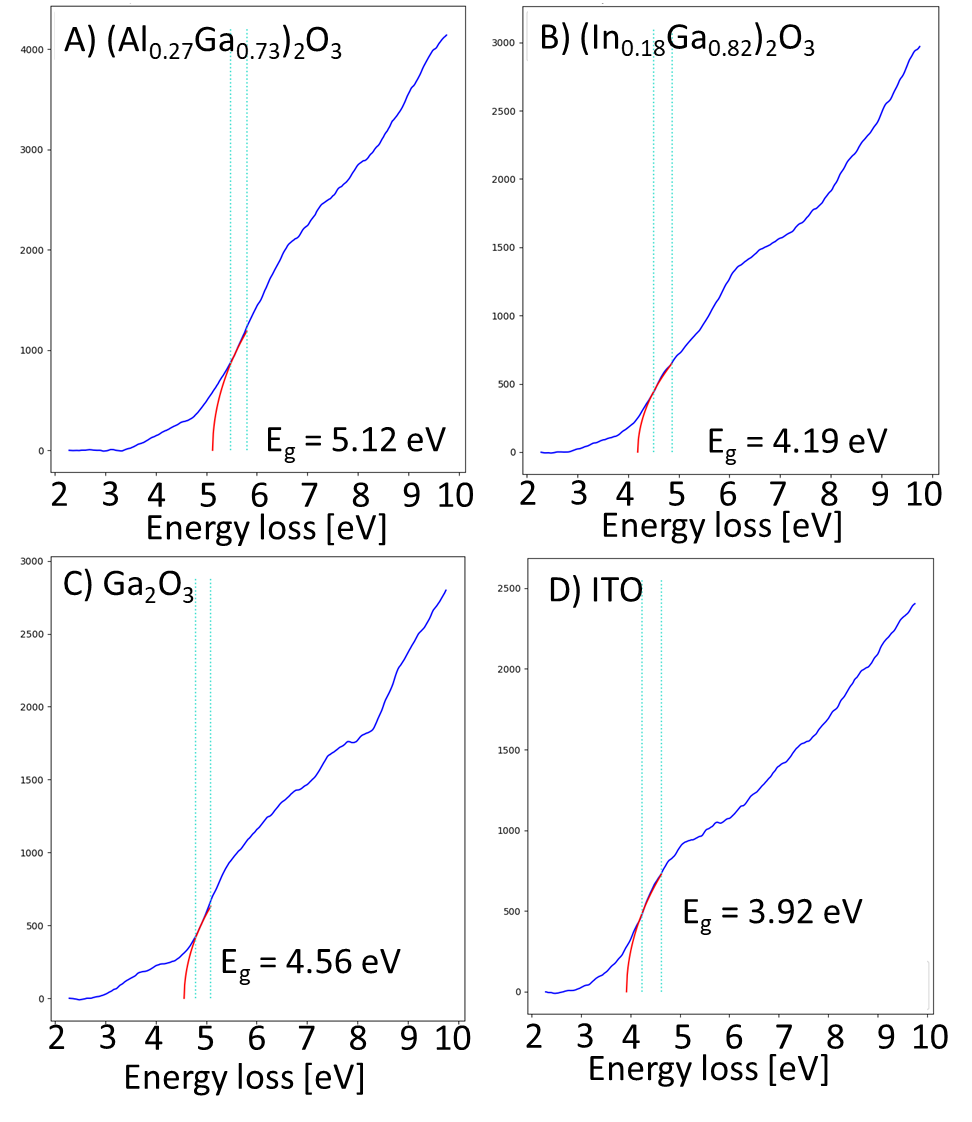}
          \caption{Fitted band gap of the A) (Al$_{0.27}$Ga$_{0.73})_2$O$_3$ layer, B) (In$_{0.18}$Ga$_{0.82})_2$O$_3$, C) Ga$_2$O$_3$, and D) ITO layer.}
          \label{figure:BG}
        \end{center}
\end{figure}

To investigate the electronic structure across the multilayered heterostructure, we performed spatially resolved band gap measurements using EELS. The STEM image of the cross-section is shown in Figure \ref{figure:TEM}A. Corresponding band gap mapping is presented in Figure \ref{figure:TEM}B, where the image intensity reflects the fitted band gap value extracted from the EELS spectrum at each pixel. Representative spectra from the four distinct layers are shown in Figure~\ref{figure:BG}, with red curves indicating the fitted models, turquoise separators denoting the fitting windows, and annotated fitting parameters. Band gap fitting was carried out using our in-house automated quantitative-based EELS analysis software (details provided in the supplementary section).

\begin{figure}
  \begin{center}
    \includegraphics[width=3.4in]{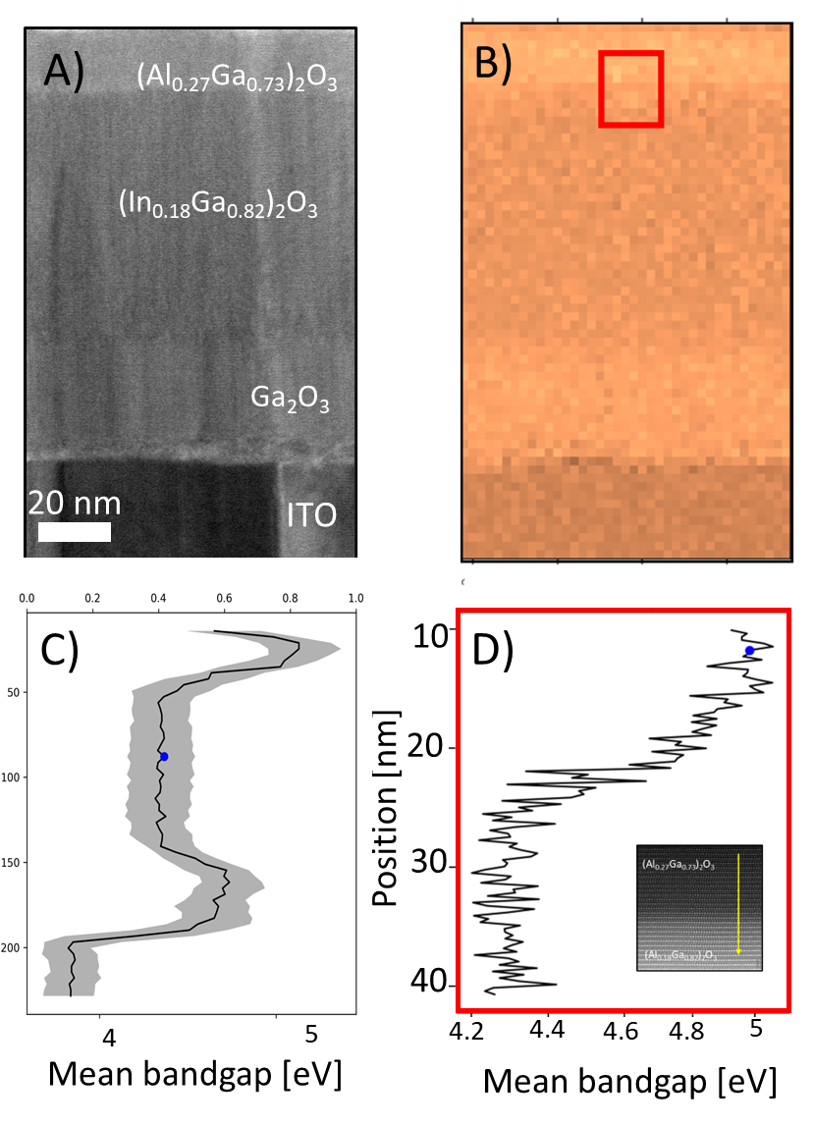}
    \label{figure:TEM}
          \caption{A) BF-STEM image of the 3-layered Ga$_2$O$_3$ on ITO substrate, B) band gap map of the same region, C) fitted mean band gap (black) and corresponding $\pm 1$ standard deviation (gray) calculated row-wise from the EELS spectrum image, and D) fitted mean band gap across the two first layers from the line scan EELS data.}
    \label{figure:TEM}
        \end{center}
\end{figure}

A multilayer sample consisting of three Ga$_2$O$_3$-based layers on an ITO substrate was investigated using STEM, as shown in Figure~\ref{figure:TEM}A. The stack comprises a pure Ga$_2$O$_3$ layer directly on the ITO, followed by (In$_{0.18}$Ga$_{0.82}$)$_2$O$_3$, and capped with (Al$_{0.27}$Ga$_{0.73}$)$_2$O$_3$. Band gap mapping across the cross-section is presented in Figure \ref{figure:TEM}B, where the image intensity corresponds to the fitted band gap extracted from EELS spectra at each pixel. Representative spectra from the four layers are shown in Figure~\ref{figure:BG}, with red curves indicating the fits, blue regions denoting the fitting windows, and annotated fitting parameters.

\begin{figure}
  \begin{center}
    \includegraphics[width=3.4in]{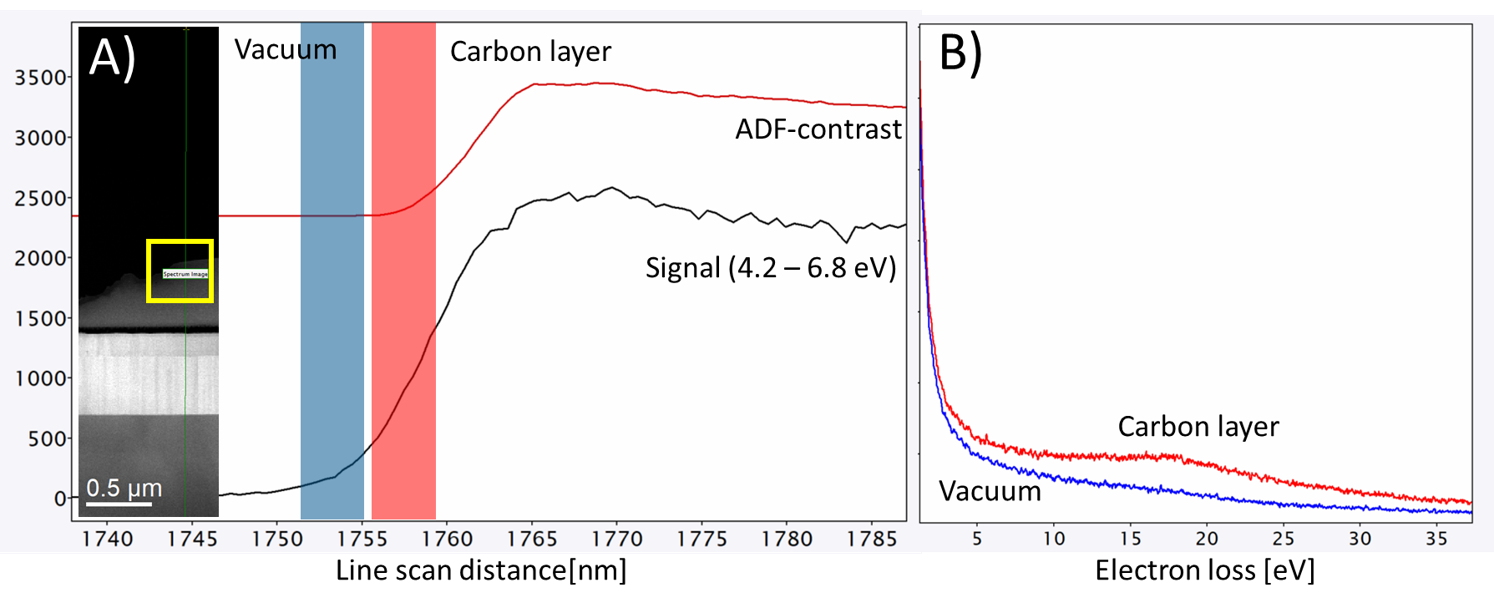}
          \caption{A) The signal from the line scan of the 3-layered Ga$_2$O$_3$ on ITO substrate, within the energy window of 4.2 eV to 6.8 eV (black curve) plotted with the intensity derived from the ADF image (red curve) near the protective carbon layer, with the ADF image as an inset. B) The EELS low loss signal from the red and blue region in (A).}
    \label{figure:EELS-C}
        \end{center}
\end{figure}

Figure \ref{figure:TEM}C displays the mean band gap along each line in the image, with gray shading indicating the standard deviation. The (Al$_{0.27}$Ga$_{0.73}$)$_2$O$_3$ layer exhibits a maximum band gap of approximately 5.1 eV, with a slight reduction near the surface and at the interface with the (In$_{0.18}$Ga$_{0.82}$)$_2$O$_3$ layer. The latter shows a band gap of 4.3 eV, while the Ga$_2$O$_3$ layer has a fitted band gap of 4.8 eV, and the ITO substrate shows 3.9 eV. As discussed in our previous work, the band gap values of the (Al$_{0.27}$Ga$_{0.73}$)$_2$O$_3$ and (In$_{0.18}$Ga$_{0.82}$)$_2$O$_3$ layers are lower than theoretical predictions, which we attribute to strain-induced lattice distortions. \cite{Thogersen:soon}

\begin{figure}
  \begin{center}
    \includegraphics[width=3.4in]{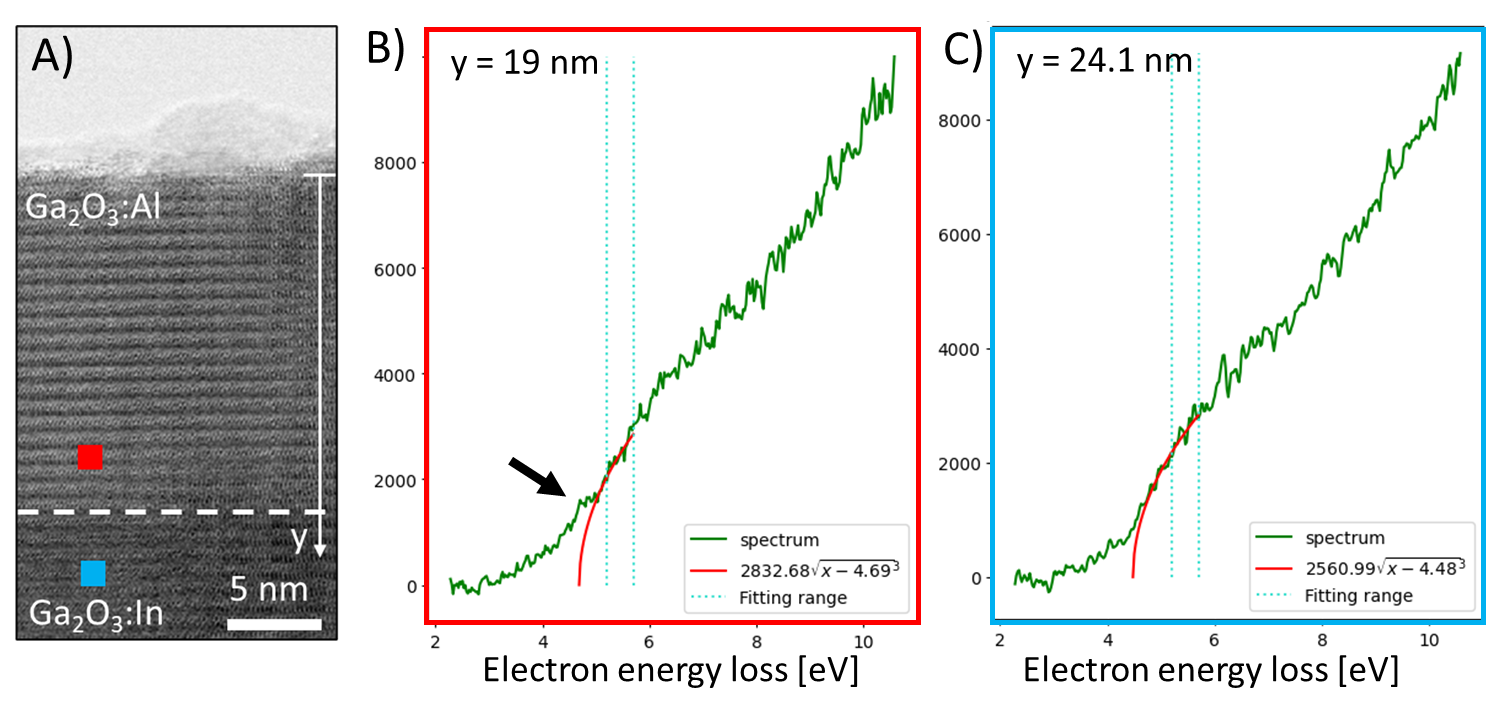}
          \caption{A) BF STEM image showing the location of the two spectra. B) Low loss EELS signal with the fitted band gap at 19 nm into the  (Al$_{0.27}$Ga$_{0.73})_2$O$_3$ layer, at the point where the band gap of both layers are visible (band gap of (In$_{0.18}$Ga$_{0.82})_2$O$_3$ shown with black arrow). C) EELS spectrum at 24.1 nm, inside the (In$_{0.18}$Ga$_{0.82})_2$O$_3$ layer. This is the point where only the band gap of this layer is visible. }
    \label{figure:eels-both}
        \end{center}
\end{figure}

A higher-resolution EELS line scan across the top two layers (Figure~\ref{figure:TEM}D) reveals a gradual band gap decrease from 5.08 eV to 4.28 eV over 10 nm. The inset in the image is a high-resolution STEM image of the ``near-perfect'' interface between the two layers, also shown in Figure \ref{figure:eels-both}B. This raises the question of whether the observed gradient reflects a true variation or is influenced by measurement limitations such as spatial resolution and inelastic delocalization. 

\begin{figure}
  \begin{center}
    \includegraphics[width=3.4in]{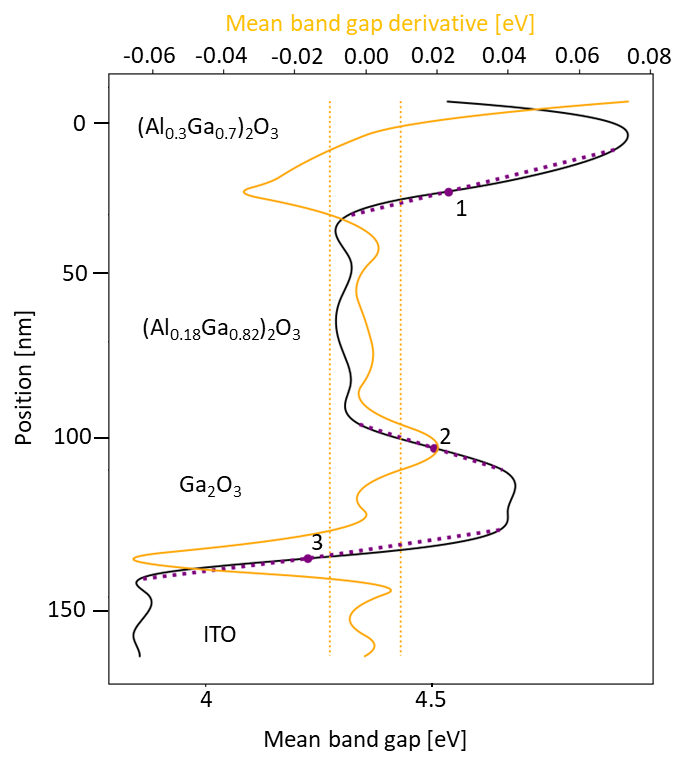}
          \caption{Smoothened line profile of the band gap (black), its derivative (orange), and flex points across the layers of the 3-layered Ga$_2$O$_3$ on ITO substrate, from EELS line scan data.}
    \label{figure:BG-line}
        \end{center}
\end{figure}

The initial approach to explore delocalization in these samples, based on our microscope parameters, involved determining the point at which a signal was observed while approaching the sample from vacuum conditions. Line scan EELS measurements were employed for this assessment. As shown in the inset of Figure \ref{figure:EELS-C}A, the scan begins 1.6 $\mu$m from the protective carbon layer. The red curve represents ADF intensity, while the black curve shows EELS signal intensity in the 4.2–6.8 eV range. The onset of detectable signal occurs 8.17 nm before reaching the carbon layer, as confirmed by low-loss spectra in Figure \ref{figure:EELS-C}B.

\begin{figure}
  \begin{center}
    \includegraphics[width=3.4in]{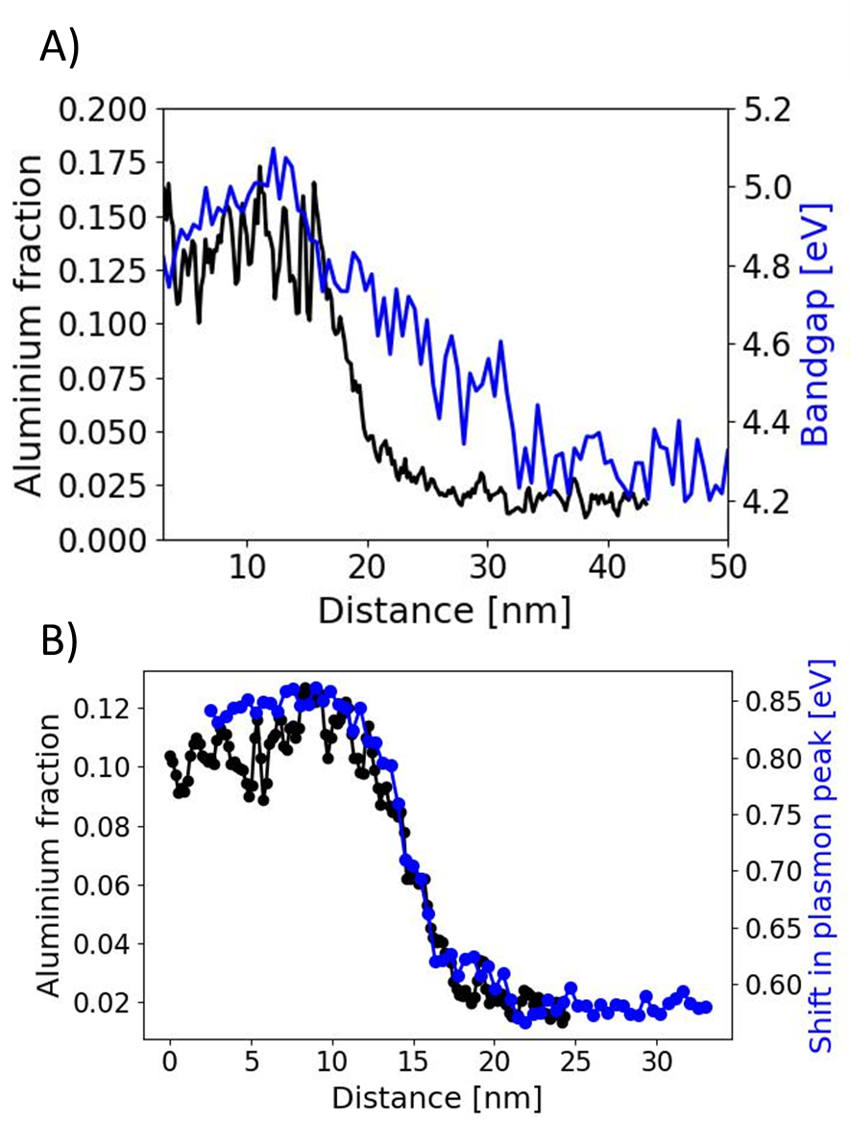}
          \caption{A) Band gap and Al fraction, and B) shift in plasmon peak and Al fraction across the (Al$_{0.27}$Ga$_{0.73})_2$O$_3$ / (In$_{0.18}$Ga$_{0.82})_2$O$_3$ interface.}
    \label{figure:BgEDS}
        \end{center}
\end{figure}

To assess whether the observed band gap gradient (\(\sim 10~\mathrm{nm}\)) could arise from probe delocalization, we also estimated the inelastic delocalization length \(L\) for our STEM-EELS conditions using Egerton's approximation for low-loss excitations \cite{Egerton_book,LowLossEELS2025}:
\[
L \approx 0.8~\mathrm{nm} \times \frac{E_0 / 100~\mathrm{kV}}{E / 10~\mathrm{eV}}
\]
where \(E_0\) is the accelerating voltage and \(E\) the energy loss. For \(E_0 = 60~\mathrm{kV}\) and \(E \approx 5~\mathrm{eV}\) (band gap region), this gives:

\[
L \approx 0.8~\mathrm{nm} \times \frac{60 / 100}{5 / 10} \approx 0.96~\mathrm{nm}.
\]

Even under conservative assumptions (small collection angle), the delocalization length remains below \(4~\mathrm{nm}\), significantly smaller than the measured \(10~\mathrm{nm}\) band gap gradient. This confirms that the gradient is not an artifact of inelastic delocalization but reflects a physical variation, likely strain-induced. 

Figures \ref{figure:eels-both}A and B provide a detailed view of the interface between (Al$_{0.27}$Ga$_{0.73}$)$_2$O$_3$ and (In$_{0.18}$Ga$_{0.82}$)$_2$O$_3$. At 19 nm into the Al-rich layer (Figure \ref{figure:eels-both}B), both band gaps are visible, while at 24.1 nm into the In-rich layer (Figure \ref{figure:eels-both}C), only the In-rich band gap remains. This suggests an interfacial region of ~5 nm where both signals overlap, likely due to delocalization.

From our exploration of the interface between Ga$_2$O$_3$ and the ITO substrate, we infer that this interface is characterized by an abrupt transition, indicating the likelihood of possessing the steepest inflection point. This presumption is grounded in the notion that abrupt interfaces often exhibit distinctive features in electronic transitions. Applying this line of reasoning to the (Al$_{0.27}$Ga$_{0.73})_2$O$_3$ / (In$_{0.18}$Ga$_{0.82})_2$O$_3$ interface, a similar abrupt nature would suggest a comparable inflection point. However, a nuanced observation emerges from Figure \ref{figure:BG-line}, where the inflection point of this interface appears to be less steep than anticipated. This intriguing finding prompts a more in-depth interpretation. It suggests the possibility that the band gap across the interface is undergoing not only a shift due to delocalization effects but also a substantial alteration. 

Variations in band gap across the interface may be partly attributed to intermixing of Al and In at the boundary. To examine this, Figure \ref{figure:BgEDS}A plots the band gap and Al concentration across the interface, alongside the shift in plasmon peak energy relative to Al concentration. The plasmon peak energy, which depends on the valence electron density of the material, shifts with Al and In in Ga$_2$O$_3$. Detailed analysis of this shift is presented in the Supplementary Information section. The figure reveals that the Al concentration decreases sharply at the interface, whereas the band gap reduction is more gradual. Interestingly, the shift in plasmon peak energy aligns closely with the Al concentration profile, suggesting that changes in measured band gap energy are not solely attributable to variations in Al composition at the interface.

\subsection{Band gap variations across nanometer thin wavy multi-layers in sample on ZnO}
\label{section:zno}

ZnO was selected as an alternative growth template to ITO primarily due to differences in lattice mismatch with $\kappa$-Ga$_2$O$_3$. While ITO provides a smooth growth surface, its cubic bixbyite structure introduces a significant lattice mismatch (approximately 7--8\%), which can induce residual strain in defect-free interfaces. In contrast, ZnO, with its wurtzite structure and closer lattice parameters along certain orientations, was expected to promote partial strain relaxation, albeit at the cost of introducing structural defects such as stacking faults. This comparison enables us to assess how strain and defect formation influence local band gap behavior in $\kappa$-Ga$_2$O$_3$ heterostructures.

\begin{figure}
  \begin{center}
    \includegraphics[width=3.4in]{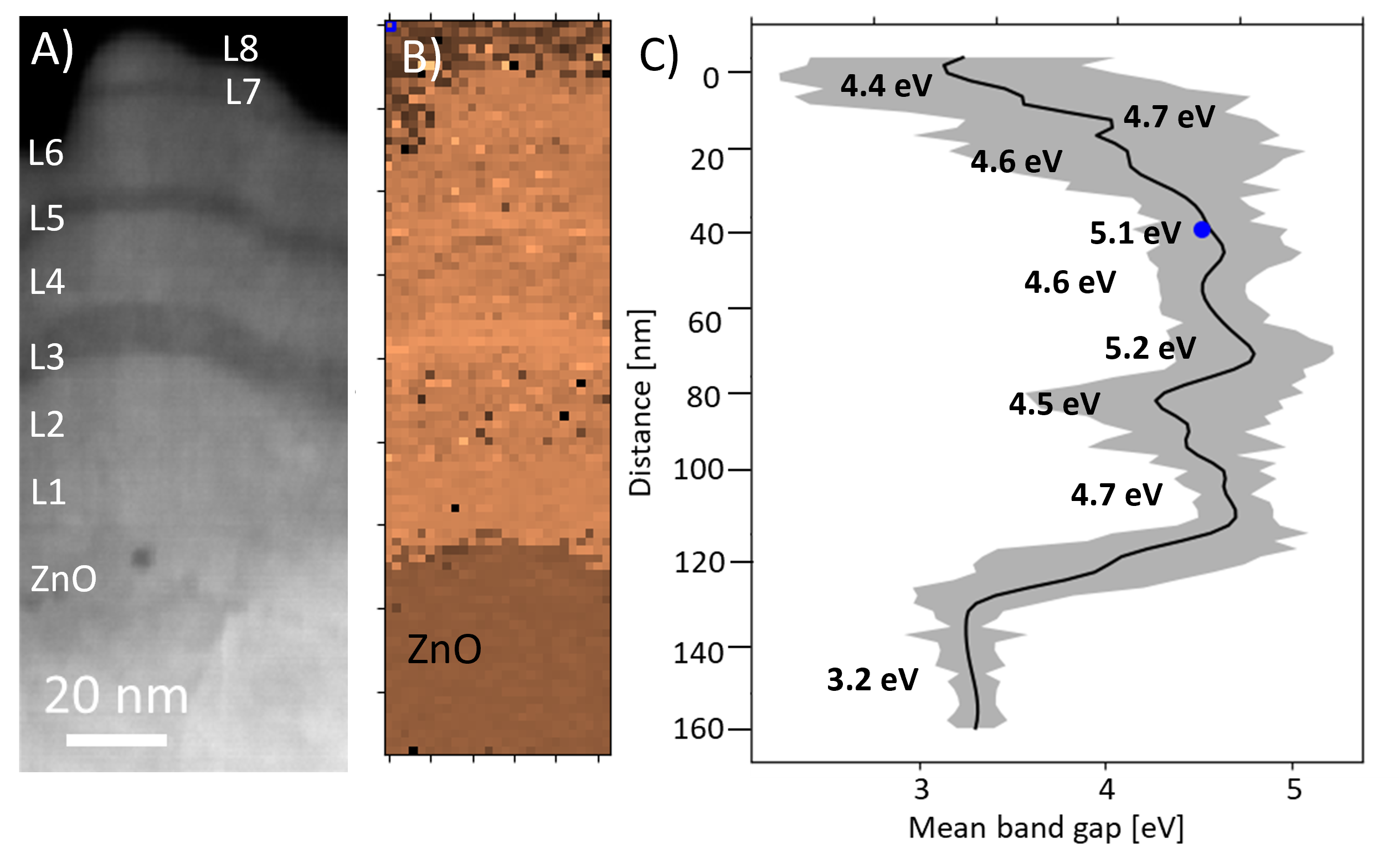}
          \caption{A) ADF-STEM image with the composition of each layer of the multilayered Ga$_2$O$_3$ on ZnO sample, B) band gap mapped image with C) the mean band gap plot, showing the variations of the band gap through the layers.}
    \label{figure:ZnO-eels}
        \end{center}
\end{figure}

The band gap variation of the nanometer thin doped, multilayered Ga$_2$O$_3$ films on ZnO substrate was then investigated. This was done to see if we could separate band gap contributions from nano-sized layers, and if there were any real or measured band gap variations when comparing thin and thick layers. An ADF image of the sample is shown in Figure \ref{figure:ZnO-eels}A, together with the band gap mapped image (B), and mean band gap curve (C). The layers' waviness and thinness allow for the detection of both Al and In in most samples using EDS. Consequently, both Al-doped and In-doped band gaps are identified when probing these layers. In Figure \ref{figure:ZnO-eels}, only the first detected band gap is shown for clarity, while Table \ref{tab:ZnO} lists both band gaps for each layer. We found that there is more variations and higher degree of difficulty to distinguish the band gap values in these samples, because the layers are thin and not planar. 

The fitted band gap values for the eight layers, along with EDS-derived Al and In fractions, layer thicknesses, and calculated band gap values, are summarized in Table~\ref{tab:ZnO}. These band gaps were measured and fitted using the same procedure as for the sample on ITO. However, in contrast to that sample, the band gaps observed here closely align with the values expected from the EDS-determined compositions. Interestingly, while defects were observed in this sample, they may have helped to relieve internal stress within the lattice. This stress relaxation could result in lattice parameters and band gap values that are closer to the expected values based on composition. In contrast, the ITO-based sample exhibited no visible defects but showed a deviation in lattice parameters, suggesting that residual strain may have been retained. There also seems to be no differences in band gaps between thick and thin layers in the sample. 

\begin{figure*}
  \begin{center}
    \includegraphics[width=7.0in]{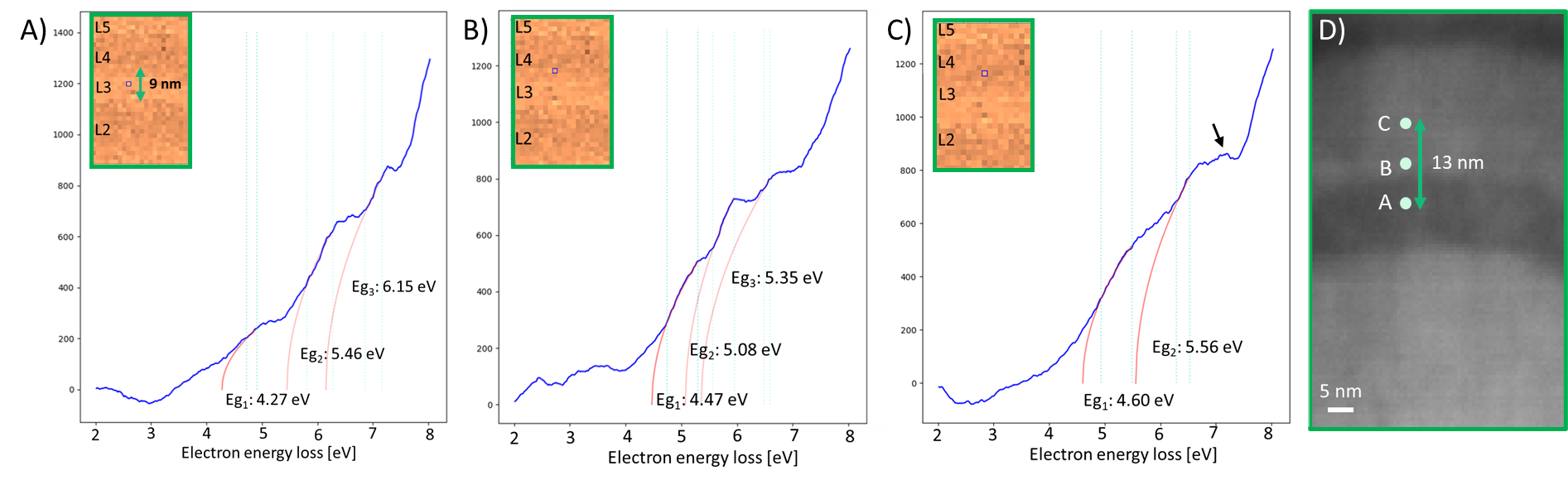}
          \caption{Low loss EELS spectra, with fitted bandgaps, from three points, A) on L3, B) in-between L3 and L4, and C) on L4. In C) the black arrow shows a small peak corresponding to the band gap at 6 eV for L3. D) shows an DF STEM image at the location of the examined layer.}
    \label{figure:ZnO-eels-BG}
        \end{center}
\end{figure*}

Figure \ref{figure:ZnO-eels-BG} presents EELS spectra from three distinct points across a thin layer, with panel A) corresponding to layer 3 (L3) in Figure \ref{figure:ZnO-eels}A, panel B) situated between L3 and L4, and C) in with L4. The spectra reveal multiple band gap excitations. In Figure \ref{figure:ZnO-eels-BG}A, excitations at 4.27 eV, 5.46 eV, and 6.15 eV are identifiable. The composition of this layer is expected to contain an aluminum fraction of 0.52 (52\% relative to total gallium content), which theoretically suggests a band gap near 6 eV. Additionally, EDS analysis indicates a presence of indium at a 0.7 fraction cation share, which could be mixed within the layer, or potentially originate from adjacent layers L2 or L4. This minimal indium fraction would likely yield a band gap around 4.57 eV, which matches our observations. 

\begin{table}
    \centering
    \begin{tabular}{c|cc|cc|cc|c}
        & \multicolumn{2}{c|}{Comp.} & \multicolumn{2}{|c|}{Calc.} & \multicolumn{2}{|c}{Exp.} &  \\
        \hline
        Layer & In & Al & Eg(In) & Eg(Al) & Eg(1) & Eg(2) &  t\\
              & \multicolumn{2}{c|}{[frac.]}  & \multicolumn{2}{c}{[eV]} & \multicolumn{2}{|c|}{[eV]} & [nm] \\
             \hline
           L8 & 0.25 & 0.25 & 4.41 & 5.43 & 4.4 & 5.4 & 20 \\ 
           L7 &      & 0.6  &      & 5.95 &     & 6.0 & 4 \\
           L6 & 0.11 & 0.33 & 4.69 & 5.59 & 4.7 & 5.6 & 18 \\
           L5 &      & 0.52 &      & 5.99 &     & 6.0 & 6 \\
           L4 & 0.11 & 0.27 & 4.69 & 5.47 & 4.7 & 5.5 & 20 \\
           L3 & 0.17 & 0.52 & 4.57 & 5.99 & 4.6 & 6.0 & 13 \\
           L2 & 0.15 & 0.25 & 4.61 & 5.43 & 4.6 & 5.4 & 20 \\
           L1 & 0    & 0.20 &      & 5.32 &     & 5.3 & 25 \\ 
           \hline
    \end{tabular}
    \caption{Composition (comp.) of In and Al in the eight layers (as fractions of the total Ga amount) in the Ga$_2$O$_3$ on ZnO sample, with the experimental (exp.) and calculated (calc.) band gap (E$_g$) values, and layer thicknesses (t).}
    \label{tab:ZnO}
\end{table}

In L4, as shown in Figure \ref{figure:ZnO-eels-BG}C, the primary doping element is indium in Ga$_2$O$_3$, but measurements also indicate a 0.27 fraction of aluminum and a 0.11 fraction of indium. These fractions result in calculated band gaps of 4.69 eV and 5.47 eV, respectively. The spectrum from L4 exhibits fitted band gaps at 4.6 eV and 5.56 eV. Intriguingly, the peak at 5.46 eV observed in Figure \ref{figure:ZnO-eels-BG}A may actually originate from L4.

Additionally, a smaller peak at 7.2 eV, marked by a black arrow, aligns with the position used to fit the 6 eV band gap in Figure \ref{figure:ZnO-eels-BG}A. The spectrum in Figure \ref{figure:ZnO-eels-BG}B, located between the two layers, features band gaps at 4.47 eV, 5.08 eV, and 5.35 eV. The first peak at 4.47 eV appears to be an amalgamation of the band gaps for L3 at 4.27 eV and L4 at 4.6 eV. Likewise, the third peak at 5.35 eV may represent a combination of the band gaps at 6.15 eV from L3 and 5.56 eV from L4.

This demonstrates the feasibility of distinguishing between distinct band gaps within the multilayers even at high resolution. At the interface, where the layer spacing is less than 7 nm, band gaps exhibiting excitations several electron volts apart can be more readily discerned, rendering their values more reliable. However, when dealing with band excitations of less than 2 eV, obtaining a reliable measurement is not as straightforward.

\section{Discussion}

Our study demonstrates that monochromated STEM-EELS combined with automated automated quantitative-based analysis enables quantitative band gap mapping across $\kappa$-Ga$_2$O$_3$ heterointerfaces with nanometer-scale precision. This approach provides direct visualization of electronic structure in regions where overlapping signals and ultra-thin layers have traditionally posed significant challenges. The ability to resolve multiple band gap excitations within sub-10 nm layers highlights the sensitivity and spatial resolution achievable with this methodology.

In the ITO-based heterostructure, measured band gaps were consistently lower than composition-based predictions, particularly in the Al$_{0.27}$Ga$_{0.73}$O$_3$ and In$_{0.18}$Ga$_{0.82}$O$_3$ layers. These deviations cannot be attributed to inelastic delocalization or intermixing alone, as the observed gradients exceed both the estimated delocalization length ($\sim$5 nm) and the fitting uncertainty ($\sim$0.1 eV). Instead, the variations are physically meaningful and strain-driven, likely amplified by the absence of interfacial defects in this high-quality structure.

GPA strain mapping provides complementary evidence for this interpretation. As shown in Figure~\ref{figure:TEMall}G, the ITO-based sample exhibits a continuous increase in compressive strain across the (In$_{0.18}$Ga$_{0.82}$)$_2$O$_3$ to (Al$_{0.27}$Ga$_{0.73}$)$_2$O$_3$ transition, correlating with the gradual band gap reduction observed by EELS. In contrast, the ZnO-based sample shows localized strain fluctuations associated with stacking faults, which likely facilitate stress relaxation and yield band gaps closer to composition-based predictions. These findings underscore that while STEM-EELS captures the electronic signature, strain analysis remains essential for interpreting deviations from theoretical expectations.

Figure~\ref{figure:BgEDS} further supports this conclusion: While the Al concentration decreases sharply at the Al$_{0.27}$Ga$_{0.73}$O$_3$/In$_{0.18}$Ga$_{0.82}$O$_3$ interface, the corresponding band gap reduction is more gradual. The close alignment between plasmon peak shifts and Al profiles reinforces the interpretation that strain, rather than intermixing or measurement artifacts, governs the band gap gradient.

The ZnO-based multilayer sample exhibited band gap values that closely matched those predicted from EDS-derived compositions. Structural defects such as stacking faults and interfacial roughness likely facilitated stress relaxation, allowing lattice parameters to approach equilibrium values and restoring band gaps to theoretical expectations. Despite the challenges posed by thin and wavy layers, multiple band gap excitations were successfully resolved, demonstrating the robustness of the automated quantitative-based EELS analysis workflow for complex multilayer systems.

The observed strain-induced band gap reduction in the ITO-based heterostructure has direct implications for device performance. In high-power or high-frequency electronic devices, such as Schottky barrier diodes (SBDs) and field-effect transistors, a reduced band gap can lead to increased leakage currents, reduced breakdown voltage, and compromised thermal stability~\cite{10.1063/5.0250729}. Conversely, the ability to engineer strain and defects to tune the band gap locally offers opportunities for band structure engineering in device design. Strain-relaxed interfaces, as observed in the ZnO-based sample, may be more suitable for applications requiring stable and predictable electronic properties.

Overall, these results establish STEM-EELS, integrated with automated automated quantitative analysis, as a powerful technique for quantitative band gap mapping in complex oxide heterostructures. When combined with GPA strain mapping and plasmon analysis, this methodology provides a comprehensive framework for studying band alignment and strain effects in next-generation wide band gap materials.

\section{Conclusion}
This work establishes a robust methodology for nanoscale band gap mapping in complex oxide heterostructures using monochromated STEM-EELS combined with a custom automated quantitative-based analysis framework. The software enables automated extraction and visualization of band gap excitations from spectral images and line scans, even in layers only a few nanometers thick. By applying this approach to $\kappa$-Ga$_2$O$_3$ heterostructures, we demonstrate that band gap variations can be resolved with sub-nanometer precision and correlated with local strain. While GPA strain mapping provides valuable context, the primary advance lies in the ability to decouple composition and electronic structure through direct EELS measurements. This methodology is broadly applicable to other wide band gap systems and offers a pathway toward high-throughput electronic charac-terization in ultramicroscopy.

\section{Acknowledgement}
We acknowledge the Norwegian Research Council for funding the GO2DEVICE project (301740) and the national infrastructures National Surface and Interface Characterisation Laboratory, NICE (195565), the Norwegian Center for Transmission Electron Microscopy, NORTEM (197405) and the Norwegian Micro- and Nano-Fabrication Facility, NorFab (295864).

\bibliography{SolOPP}

\end{document}